\documentclass[twocolumn,showpacs,pra,amsmath,amssymb,eqsecnum]{revtex4}
\usepackage[T1]{fontenc}
\usepackage[latin9]{inputenc}
\usepackage{graphicx}
\usepackage{esint}



\usepackage[T1]{fontenc}
\usepackage[latin9]{inputenc}

\usepackage{esint}
\usepackage{epsfig}
\usepackage{amsmath}
\usepackage{latexsym}

\usepackage{amssymb,amsmath}

\usepackage{graphicx}% Include figure files
\usepackage{dcolumn}% Align table columns on decimal point
\usepackage{bm }% bold math
\usepackage{epstopdf}
\begin{document}

\title{The electric field of a point charge in a spherical inclusion structure}

\author{Asaf Farhi}
 \email{asaffarhi@post.tau.ac.il}
\author{David J. Bergman}
\email{bergman@post.tau.ac.il}

\affiliation{
Raymond and Beverly Sackler School of Physics and Astronomy,
Faculty of Exact Sciences,
Tel Aviv University, IL-69978 Tel Aviv, Israel
%, and Department of Physics,
%The Ohio State University, Columbus, OH 43210-1117, USA
}
\begin{abstract}
A point charge in the presence of a metallic nanoshpere is a fundamental setup, which has implications for Raman scattering, enhancement of spontaneous emission
of a molecule by an antenna, sensing, and modeling a metallic tip in proximity to a nanoparticle. Here, 
we analytically expand the electric field of a point charge in an $\epsilon_{2}$ host medium
in the presence of an $\epsilon_{1}$ sphere using the sphere eigenstates, where $\epsilon_1$ and $\epsilon_2$ can take any values.
Only the $m=0$ spherical harmonics are employed in the expansion
and the calculation of the potential and the electric field is very simple. 
The electric field is strongly enhanced when  $\epsilon_{1}/\epsilon_2$  is close to 
an  $(\epsilon_{1}/\epsilon_2)_l$ eigenvalue of a dominant mode, which is determined by the point charge location and the measurement point. An electric
field exists inside the sphere when $\epsilon_{1}/\epsilon_2$ is close to a $(\epsilon_{1}/\epsilon_2)_l$ resonance even when $\epsilon_1$ is a conductor. Low order modes generate an electric field far away from the interface, where the $l=1$ mode with a resonance at $\epsilon_1=-2\epsilon_2$ generates a field at the sphere center. The high order modes which are associated with high spatial frequencies become more dominant when the point charge approaches the sphere surface or when the physical parameters are close the high order modes resonances. When  $\epsilon_{1}/\epsilon_2$ is smaller or larger than the eigenvalues of the dominant modes, the modes interfere constructively and generate a strong signal at an angular direction equal to that of the source. The spectral information at the sphere surface may be utilized to calculate the point charge location without knowing its magnitude.
 %When scaling down the setup by a factor of $b$ the intensity of the scattered electric field scales up by a factor of $b^2$.
\end{abstract}
\date{\today}
\maketitle

\section{Introduction}

The electrostatic potential of a point charge in proximity to a conducting
sphere was calculated analytically long ago \cite{jackson1975electrodynamics}. This calculation
assumes a constant potential on the sphere envelope and uses the method
of images to construct a potential outside the sphere. Scattering
eigenstates of Maxwell's equations have been exploited to calculate
the electric field in electrostatics \cite{bergman1979dielectricJPC,bergman1979dielectricPRB,bergman2014perfect,Farhi2014exact,LectNotes83}
and in electrodynamics \cite{bergman1980theory,farhi2016electromagnetic}.
Recently, a procedure to treat current sources using the electrodmagnetic spectral 
expansion has been introduced \cite{farhi2016electromagnetic}. Here,
we introduce a procedure to treat charge sources using the electrostatic eigenstate
expansion. In addition, we show that when the system is close to a resonance strong electric field exists inside the sphere even if it is a conductor.

Potential applications are enhancement of spontaneous emission
of a molecule by an antenna \cite{eggleston2015optical} in the quasistatic regime, modeling a tip in proximity to a nanosphere, near field imaging, sensing, and Raman spectroscopy.
In particular, enhancement of Raman scattering and spontaneous emission of a molecule become more dominant when the field intensity is higher, which can be obtained when the system is close to a resonance.
Near field imaging exploits evanescent waves to generate an image with resolution which is better than the diffraction limit. In this technique a 2D image is generated by scanning the surface with a scattering tip. We show that the spectral information of the electric field can be utilized to calculate the point charge location when it is not at the sphere surface, which we define as the detector.

In Sec.\ II we present the theory and introduce a procedure to treat
charge sources. In Sec.\ III we describe how we can obtain the point charge location from the spectral content of the electric field on the sphere surface. In Sec.\ IV we present the potential and the electric
field for permittivity values which are close to
the resonances of the dominant eigenstates. In Sec.\ V we discuss our results
and potential applications.

\section{Theory}

In the quasistatic regime Maxwell's equations reduce to Poisson's equation for the electric potential $\psi$
\begin{equation}
\nabla\cdot\left(\epsilon\nabla\psi\right)=-4\pi\rho.
\end{equation}

By expressing the permittivity using the step functions $\theta_{1},\theta_{2}$
of the the $\epsilon_{1}$ and $\epsilon_{2}$ media [$\theta_i(\mathbf{r})=1$ when $\epsilon(\mathbf{r})=\epsilon_i,$ otherwise $\theta_i(\mathbf{r})=0$] we write
\begin{align}
\nabla\cdot\left(\left(\epsilon_{1}\theta_{1}+\epsilon_{2}\theta_{2}\right)\nabla\psi\right)&=-4\pi\rho,\nonumber \\
\nabla^{2}\psi=-4\pi\rho+u\nabla\cdot\left(\theta_{1}\nabla\psi\right)&,\,\,u\equiv1-\frac{\epsilon_{1}}{\epsilon_{2}}.
\end{align}
This is transformed to \cite{bergman1979dielectricJPC}
\begin{equation}
\label{eq:main}
\psi=\psi_{0}+u\Gamma\psi, 
\end{equation}
where
\begin{align}
\Gamma\psi&=\int dV'\theta_{1}\left(\mathbf{r}'\right)\nabla'G_{0}\left(\mathbf{r},\mathbf{r}'\right)\cdot\nabla'\psi\left(\mathbf{r}'\right),\nonumber \\
&G_{0}=\frac{1}{4\pi\left|\mathbf{r}-\mathbf{r}'\right|},\,\,\psi_{0}=\frac{q}{\epsilon_2\left|\mathbf{r}-\mathbf{r}_{0}\right|},
\end{align}
and $\mathbf{r}_{0}$ is the point charge position.

The eigenstates satisfy Eq.\ (\ref{eq:main}) when there is no source, namely
\begin{equation}
\label{eq:eigenvalue}
s_{n}\psi_n=\Gamma\psi_n,\,\,\,\,\,\frac{1}{s_{n}}\equiv u_{n}=1-\frac{\epsilon_{1n}}{\epsilon_{2}}.
\end{equation}

\noindent
By defining the scalar product
\begin{equation}
\label{eq:inner}
\left\langle \psi|\phi\right\rangle \equiv\int dV\theta_{1}\nabla\psi^{*}\cdot\nabla\phi,
\end{equation}
$\Gamma$ becomes an Hermitian operator and therefore it has a complete
set of eigenfunctions. We insert the unity operator in Eq.\ (\ref{eq:main}) and arrive at
\begin{equation}
\psi=\psi_{0}+\sum_{n}\frac{s_{n}}{s-s_{n}}\left\langle {\psi}_{n}|\psi_{0}\right\rangle {\psi}_{n},\,\,s\equiv 1/u
\end{equation}
where ${ \psi}_{n}$ are the normalized eigenstates. 

\noindent
The sphere eigenstates are \cite{bergman1979dielectricJPC}
\begin{equation}
\label{eq:eigenstates}
 {\psi}_{n}\equiv\psi_{lm}\left(\mathbf{r}\right)=\frac{Y_{lm}\left(\Omega\right)}{\left(la\right)^{1/2}}\cdot\left\{ \begin{array}{cc}
\left(\frac{r}{a}\right)^{l} & r<a\\
\left(\frac{a}{r}\right)^{l+1} & r>a
\end{array}\right.,
\end{equation}
where $a$ is the sphere radius, $Y_{lm}$ are the spherical harmonics,
and the eigenvalues are 
\begin{equation}
\label{eq:eignvalues}
\epsilon_{1l}=-\epsilon_{2}\frac{l+1}{l},\,\,\,\, s_{lm}\equiv s_{l}=\frac{l}{2l+1}.
\end{equation}
Clearly at the $l\rightarrow\infty$ limit, $s_{l}\rightarrow 1/2.$ Thus, for
a choice of $s\approx1/2$ the high order modes make a large contribution
to the potential \cite{bergman1979dielectricJPC,bergman1979dielectricPRB,LectNotes83}.

Now we proceed to calculate the scalar product $\left\langle \psi_{lm}|\psi_{0}\right\rangle .$
A direct calculation according to Eq.\ (\ref{eq:inner}) is difficult due to the
fact that $\psi_{0}$ is not trivially expressed as a function of $r.$ We therefore
exploit the fact that $\psi_{0}\left(\mathbf{r}\right)=4\pi /\epsilon_2 \int G\left(\mathbf{r},\mathbf{r}'\right)\rho(\mathbf{r}') dV'$
and use Eq.\ (\ref{eq:eigenvalue}) to obtain
\begin{align}
\left\langle \psi_{lm}|\psi_{0}\right\rangle &=\frac{4\pi}{\epsilon_{2}}\intop\int\theta_{1}\nabla\psi_{lm}^{*}\cdot\nabla G\left(\mathbf{r},\mathbf{r'}\right)\rho\left(\mathbf{r}'\right)dV'dV \nonumber \\
&=\frac{4\pi }{\epsilon_{2}}s_l\intop\psi_{lm}^{*}\left(\mathbf{r}'\right)\rho\left(\mathbf{r}'\right)dV' \nonumber \\
&=\frac{4\pi q }{\epsilon_{2}}s_l\psi_{lm}^{*}\left(\mathbf{r}_{0}\right),
\label{eq:inner2}
\end{align}
where we assumed a point charge $\rho=q\delta^3(\mathbf{r}-\mathbf{r}_0)$.
We finally get
\begin{equation}
\label{eq:expansion}
\psi\left(\mathbf{r}\right)=\psi_{0}\left(\mathbf{r}\right)+\frac{4\pi q}{\epsilon_{2}}\sum_{l,m}\frac{s_{l}^{2}}{s-s_{l}}\psi_{lm}^{*}\left(\mathbf{r}_{0}\right)\psi_{lm}\left(\mathbf{r}\right).
\end{equation}
It can readily be seen from Eqs.  (\ref{eq:eigenstates}) and (\ref{eq:expansion}) that as the point charge approaches the sphere interface $\psi_{lm}^{*}\left(\mathbf{r}_{0}\right)$ of the high order modes becomes non-negligible 
 and they become more dominant in the expansion. In addition, low order modes decay more slowly away from the interface and can therefore generate fields far away from the interface. 
% Finally, we see from from Eqs.  (\ref{eq:eigenstates}) and (\ref{eq:expansion}) that when scaling down the system by a factor of $b,$ $r/a$ and $z_0/a$ remain fixed but due to the $1/\sqrt {a},$  $\psi_{lm}^{*}\left(\mathbf{r}_{0}\right)\psi_{lm}\left(\mathbf{r}\right)$ scale down by a factor of $b.$  

The ratio $\epsilon_1/\epsilon_2$ can be chosen to enhance a contribution to the electric field of one or more modes. We can therefore decompose each term in the sum in Eq.\ (\ref{eq:expansion}) into $\left(4\pi q/\epsilon_{2}\right)s_{l}^{2}\psi_{lm}^{*}\left(\mathbf{r}_{0}\right)\psi_{lm}\left(\mathbf{r}\right),$ which does not depend on $s,$ and $1/(s-s_l)$ which is determined by the distance between physical $s$ and an eigenvalue $s_l.$

For a point charge at $\mathbf{r}_0=z_{0}\hat{\mathbf{z}},$ $\psi_{lm}^{*}\left(\mathbf{r}_0\right)=\psi_{lm}\left(\mathbf{r}_0\right)$ and $\,\psi_{lm}\left(\mathbf{r}_0\right)\neq 0$
 only when $m=0.$ Thus, $\psi\left(\mathbf{r}\right)$ is independent of the azimuthal angle $\phi$ and the sum in the last equation is considerably simplified. In addition, it can be seen that when the ratio $\epsilon_{1}/\epsilon_{2}$ is fixed, $\psi\left(\mathbf{r}\right)/\psi_{0}\left(\mathbf{r}\right)$ is also fixed since $\epsilon_{2}$ cancels out. Therefore,  the relative effect of a sphere inclusion on the potential and the electric field does not change when keeping this ratio fixed, even when $\epsilon_1$ is large. For example, the $l=1$ resonance occurs when $\epsilon_1\approx -2\epsilon_2$. Since $\mathrm{Im}\left(\epsilon_2\right)\approx-\mathrm{Im}\left(\epsilon_1\right)/2$ if $\epsilon_1$ has small dissipation, $\epsilon_2$ with smaller gain is required for the resonance and therefore $\epsilon_2$ with small dissipation will still be close to the resonance. In addition, this mode extends far from the interface and  generates fields far from the interface. When down scaling the system by a factor $b$ we get that $\left|\mathbf{E}\right|^{2}$ increases by a factor of $b^4$ as is the case for a point charge in a uniform medium. 

To verify our result in Eq.\ (\ref{eq:expansion}) we placed a point charge at $\mathbf{r}_0=z_{0}\hat{\mathbf{z}}$ and took the $\epsilon_{1}\rightarrow\infty$
limit, assuming $\epsilon_2$ is finite. We then summed a geometric series to obtain the known textbook result for $\mathbf{r}$ on the positive $z$ axis 
\begin{equation}
\psi\left(\mathbf{r}\right)=\psi_{0}\left(\mathbf{r}\right)-\frac{qa/z_{0}}{|r-a^{2}/z_{0}|}.
\end{equation}
% where $z_{0}$ is the charge position on the $z$ axis.
\noindent
The electric field can be written as follows
\begin{equation}
\label{eq:expansion_E}
\mathbf{E}\left(\mathbf{r}\right)=-\nabla\psi_{0}\left(\mathbf{r}\right)-\mathbf{E}_{\mathrm{scat}},
\end{equation}
where
$$\mathbf{E}_{\mathrm{scat}}\equiv -\frac{4\pi q}{\epsilon_{2}}\sum_{l}\frac{s_{l}^{2}}{s-s_{l}}\psi_{lm}^{*}\left(\mathbf{r}_{0}\right)\nabla\psi_{lm}\left(\mathbf{r}\right),$$
\begin{widetext}
\begin{equation}
\label{eq:electric_field}
\nabla\psi_{lm}\left(\mathbf{r}\right)=\mathbf{e}_{r}Y_{lm}\frac{\partial f_{l}\left(r\right)}{\partial r}+\mathbf{e}_{\phi}\frac{f_{l}\left(r\right)}{r\sin\theta}imY_{lm}+\mathbf{e}_{\theta}\frac{f_{l}\left(r\right)}{r}\frac{\partial Y_{l,m}}{\partial\theta},
 \end{equation}
 \[
f_{l}\left(r\right)=\frac{1}{\left(la\right)^{1/2}}\left\{ \begin{array}{cc}
\left(\frac{r}{a}\right)^{l} & r<a\\
\left(\frac{a}{r}\right)^{l+1} & r>a
\end{array}\right.,\,\,\,\,\,\,\,\,\,\,\,\,\frac{\partial f_{l}\left(r\right)}{\partial r}=\frac{1}{\left(la\right)^{1/2}}\left\{ \begin{array}{cc}
l\left(\frac{r}{a}\right)^{l}\frac{1}{r} & r<a\\
-\left(l+1\right)\left(\frac{a}{r}\right)^{l+1}\frac{1}{r} & r>a
\end{array}\right.,
\]
and where $\partial\psi_{lm}/\partial\theta$ can be written as \cite{edmonds1957angular}
\begin{equation}
\frac{\partial\psi_{lm}}{\partial\theta}=\mathbf{e}_{\theta}\frac{f_{l}\left(r\right)}{r\sin\theta}\left[\frac{l\left(l+1\right)}{\left[\left(2l+1\right)\left(2l+3\right)\right]^{1/2}}Y_{l+1,m}-\frac{l\left(l-1\right)}{\left[\left(2l-1\right)\left(2l+1\right)\right]^{1/2}}Y_{l-1,m}\right].
\end{equation}
\end{widetext}

Note that the field of the $l=1$ mode does not vanish at the origin, and that Legendre polynomials satisfy $p_l(x=1)=1,$  $Y_{l,m=0}\left(\theta=0\right)=1.$ In addition, Legendre polynomials are even or odd functions and therefore $Y_{l ,m=0}\left(\theta=\pi\right)=\pm1.$ Hence, when $s$ is very close to a resonance a dominant mode is excited and the intensity peaks at both $\theta=0$ and $\theta=\pi.$ 
 The $l$ components $E_{\mathrm{scat},\, r,l}$ at $\theta=0$ have a positive sign for $s_l<s$ and a negative sign for $s_l>s.$ 
   The $l$ components $E_{\mathrm{scat},\, r,l}$ at $\theta=\pi$ have alternating signs but at the transition between $s_l<s$ and $s_{l+1}>s$ since the coefficient $1/(s-s_l)$ also changes sign $E_{\mathrm{scat},\, r,l}$ and  $E_{\mathrm{scat},\, r,l+1}$ have the same sign. Thus, when $s\approx (s_l+s_{l+1})/2$ the dominant $l$ and $l+1$ modes will interfere destructively at $\theta=0$  and interfere constructively at  $\theta=\pi.$ When $s<s_l$ for every $l$ which corresponds to $\epsilon_{1}\gtrsim-2\epsilon_{2},$ all  $E_{\mathrm{scat},\, r,l}$ at $\theta=0$ have the same sign and they interfere constructively to generate a strong signal. In this case the low order modes which extend far from the sphere surface are strongly enhanced. Similarly, when $s>s_l$ for every $l$ which corresponds to $\epsilon_{1}\lesssim-\epsilon_{2},$ all  $E_{\mathrm{scat},\, r,l}$ at $\theta=0$ interfere constructively and a strong signal is expected there. In this case the high order modes which are associated with high spatial frequencies are strongly enhanced. When $s>s_l$ or $s<s_l$ for every $l$ the signs of $E_{\mathrm{scat},\, r,l}$ alternate at $\theta=\pi$ and a relatively weak signal is expected there.

\section{Calculating the point charge location from the spectral content of the electric field}

In the far field, a point in the object is mapped into a point in the image due to constructive interference, enabling 3D imaging. 
Near field imaging exploits evanescent waves and achieves resolutions better than the diffraction limit. However, measuring an electric field in the near field region produced by a point source which is not very close to the detector is usually difficult. This is since the modes decay exponentially with distance and since there can be orders of magnitude differences among electric field intensities produced by point sources at different distances from the detector. When we are close to a resonance, the local physical field is enhanced and there is a significant field also due to point sources that are not very close to the detector (e.g., at the sphere surface). Thus, high order components of the electric field can be detected. For a single point charge source, which we will treat as the object, the image field itensity will be maximal at an angular direction equal to that of the source and at the reflected direction vis-a-vis the spherical surface (see Sec. II).

We start by calculating the field at the sphere surface and requiring full retrieval of an $l$ mode of the electric field.
We expand $\psi_{0}$ inside the sphere, where there are no sources, using the unity operator. We then take the gradient to obtain the following expression for the electric field which is valid inside the sphere
  \begin{align}
  \mathbf{E}_{\mathrm{inside}}&=-\sum_{l,m=0}\left[\left\langle \psi_{l,m}|\psi_{0}\right\rangle \nabla\psi_{lm}+\frac{s_{l}}{s-s_{l}}\left\langle \psi_{lm}|\psi_{0}\right\rangle \nabla\psi_{lm}\right]\nonumber\\
& =-\sum_{l,m=0}\frac{s}{s-s_{l}}\left\langle \psi_{lm}|\psi_{0}\right\rangle \nabla\psi_{lm}.
\label{eq:field_inside}
\end{align}
We calculate from this expression the electric field at $r=a^{+},$ i.e. just outside the sphere, using continuity conditions (note that the rhs is taken at $\mathbf{r}=a^-,$ not $\mathbf{r}=a^+)$ 
\begin{align}
&\mathbf{E}\left(r=a^{+},\theta\right)\nonumber\\
&=-\sum_{l,m=0}\frac{s}{s-s_{l}}\left\langle \psi_{lm}|\psi_{0}\right\rangle \left(\frac{\epsilon_{1}}{\epsilon_{2}}\frac{\partial\psi_{lm}}{\partial r}\hat{\mathbf{r}}+\frac{1}{r}\frac{\partial\psi_{lm}}{\partial\theta}\hat{\boldsymbol{\theta}}\right)_{r=a^{-}}. 
\label{eq:field_surface}
 \end{align} 
\noindent
%where we have  to calculate $\mathbf{E}\left(r=a^+\right)$ from $\mathbf{E}(r=a^-).$   
%From Eq. (\ref{eq:expansion_E}) it can be seen that the radial components of $\mathbf{E}_{\mathrm{scat}}, \,E_{\mathrm{scat},\, r}$ are dire  of $\mathbf{E}_{0}$ and $\mathbf{E}_{\mathrm{scat}},$  $E_{0,r}$ and $E_{\mathrm{scat},\, r}$ have opposite signs at $\mathbf{r}=a^{+}\hat{z}$
%  and the same sign at $\mathbf{r}=-a^{+}\hat{z}.$ Thus, there is a constructive interference at $\mathbf{r}=-a^{+}\hat{z}$
%  and a destructive interference at $\mathbf{r}=a^{+}\hat{z}.$
%  When $s\simeq s_{l}$ there is a dominant mode in $E_{\mathrm{scat},\, r}.$

%An $l$ component of $E_{\mathrm{scat},\, r},$  and $E_{0,r}$
% at $\mathbf{r}=a^{+}\hat{z}$ have the same sign for $s<s_{l}$
% and a different sign for $s>s_{l},$ assuming that $\epsilon_{1}$
%  and $\epsilon_{2}$ have different signs.

The magnitudes of the high order modes in the expansion of a point charge field in a uniform medium become smaller as the point charge is farther from the sphere surface (see the expansion of $\psi_0$ in Eqs. (\ref{eq:field_inside}) and (\ref{eq:field_surface})). We treat $\mathbf{r}=a^+\hat{\mathbf{z}}$ as the measurement point and require that for a given mode the electric field enhancement due to the presence of the sphere will compensate for the decay of the field due to the distance between the point charge source and the measurement point. We therefore require that an $l$ component of the electric field of a point charge at $\mathbf{r}_{0}=z_{0}\hat{\mathbf{z}}$
  measured at the sphere surface at $\mathbf{r}=a^{+}\hat{\mathbf{z}}$ and an $l$ component of the electric field of a point charge in a uniform medium at $\mathbf{r}_0=a^{+}\hat{\mathbf{z}}$ and a measurement point also at $\mathbf{r}=a^{+}\hat{\mathbf{z}}$ will be equal
\begin{align}
&\frac{E_{r,l,\,\mathrm{sphere\, setup}}\left(\mathbf{r}=a^{+}\hat{\mathbf{z}},\,\mathbf{r}_{0}=z_{0}\hat{\mathbf{z}}\right)}{E_{r,l,\,\mathrm{uniform\, medium}}\left(\mathbf{r}=a^{+}\hat{\mathbf{z}},\mathrm{\mathbf{r}_{0}=a^{+}\hat{\mathbf{z}}}\right)}
 \nonumber\\
&=\frac{\frac{\epsilon_{2}}{\epsilon_{1}}\frac{ss_{l}}{s-s_{l}}\psi_{lm}\left(\mathbf{r}_{0}=z_{0}\hat{\mathbf{z}}\right)\left.\frac{\partial\psi_{lm}}{\partial r}\right|_{\mathbf{r}=a^{-}\hat{\mathbf{z}}}}{s_{l}\psi_{lm}\left(\mathbf{r}=a^{+}\hat{\mathbf{z}}\right)\left.\frac{\partial\psi_{lm}}{\partial r}\right|_{\mathbf{r}=a^{-}\hat{\mathbf{z}}}}\nonumber\\
&=\frac{\epsilon_{2}}{\epsilon_{1}}\frac{s}{s-s_{l}}\left(a/z_{0}\right)^{l+1}\simeq1,
 \end{align} 
 where we have used $E_{r,l}\left(\mathbf{r}=a^{-}\hat{\mathbf{z}}\right)=E_{r,l}\left(\mathbf{r}=a^{+}\hat{\mathbf{z}}\right)$ for a point charge in a uniform medium.
\noindent
Assuming $s\simeq1/2,\, s_{l}-s\simeq0.0025$ we obtain
$$\left(a/z_{0}\right)^{l+1}\simeq2\left(s_{l}-s\right)\simeq0.005,$$
 and for $l=10$ we get
$$z_{0}/a\simeq1.62.$$
\noindent
This means that if we assume $\epsilon_{1}\simeq-\epsilon_{2},\, s-s_{l}\simeq \left(\epsilon_{1}-\epsilon_{1l}\right)/(4\epsilon_2)$ and for $\epsilon_2=1.5,\,\epsilon_1-\epsilon_{1l}\simeq0.015$ and  $\ensuremath{z_{0}/a\lesssim1.62},$ the $l=10$ mode magnitude is equal to or higher than its magnitude when measuring the electric field at the point charge location (uniform medium). The angular half width of this mode near $\theta=\pi$ calculated using the $l=10$ Legendre polynomial is 0.14 rad which translates to 4nm for a sphere with a radius of 30nm. Note that if $\epsilon_1$ has dissipation we can get closer to a real $s_l$ by using $\epsilon_2$ with gain \cite{Farhi2014exact}.  

We now calculate the point charge location using the spectral content of the electric field on the sphere surface.
The electric field of a point charge at the sphere surface is composed of modes with magnitudes which depend on the point charge location. 
Thus, the spectral information of the electric field is affected by the point charge location. If $s\approxeq s_{l},$ the electric field is dominated by this $l$ mode. Alternatively, if the radial component of the electric field on the sphere surface can be measured then by using a spherical harmonics transform defined by
  \begin{equation}
  F\left(l,m\right)=\int E_{r}Y_{lm}^{*}d\Omega,
 \end{equation}
 we can obtain the spectral content of an $l,m$ mode in the expansion of the physical electric field. Note that this transform gives the spectral content since $\int Y_{l'm'}Y_{lm}^{*}d\Omega=\delta_{ll'}\delta_{mm'}$ and $E_{r,lm}$ has a $Y_{l,m}$ associated with it. To perform the transform we need to choose a coordinate system so that $\theta=0$ points to the point charge location. Since the maximal intensity is always at $\theta=0,\pi$ we must choose between them to define $\theta=0$ according to the $s$ value (see discussion above) or by knowing in which half-space the point charge is located.
 The ratio between the magnitudes of the $l_1$ and $l_2$ components of the electric field of a point charge located at $\mathbf{r}_0=z_{0}\hat{\mathbf{z}}$ 
 is 
 \begin{equation}
 \frac{F\left(l_{1},m=0\right)}{F\left(l_{2},m=0\right)}=\frac{l_{1}+1}{l_{2}+1}\frac{l_{2}}{l_{1}}\frac{s_{l_{1}}}{s_{l_{2}}}\frac{s-s_{l_{2}}}{s-s_{l_{1}}}a^{l_{1}-l_{2}}z_{0}^{l_{2}-l_{1}}.
 \end{equation}
 Thus, from this ratio we can calculate the point charge location $z_0.$ Now using $z_0$ it is straightforward to calculate $q$ from any $F(l,m)$ component.
%The ratio between the magnitudes of an $l$ component of the electric fields of point charges $q_{1}$
% and $q_{2}$ located at $z_{01}\hat{z}$ and $z_{02}\hat{z},$ respectively,
% is $\psi_{lm=0}\left(\mathbf{r}_{01}\right)/\psi_{lm=0}\left(\mathbf{r}_{02}\right)
% =q_{1}/q_{2}\cdot\left(z_{02}/z_{01}\right)^{l+1}.$ Hence, if we know the point charge locations $z_{01}\hat{z}$ and $z_{02}\hat{z}$ we can deduce the r charge magnitudes $q_1$ and $q_2$ and the magnitudes of an $l$ component of the electric fields of two point charges, we can deduce the ratio between the point charge locations.
 In order for the $l$ mode fields of two point charges $q_{1}$ and $q_{2}$ located at $z_{01}\hat{\mathbf{z}}$ and $z_{02}\hat{\mathbf{z}},$ respectively, to be comparable in magnitude we can require $0.1\lesssim q_{1}/q_{2}\cdot\left(z_{02}/z_{01}\right)^{l+1}\lesssim10.$ For example, for the $l=10$ mode assuming $q_1=q_2$ we obtain that for comparable field intensities we must have $0.9 \lesssim z_{01}/z_{02}\lesssim 1.11.$  Thus, objects in a range of 3nm along $r$ for a sphere with a radius of 30nm produce comparable field intensities at the sphere surface. 
% When dividing the ratio between an $l_1$ mode electric fields of two point charges by the ratio between an $l_2$ mode electric fields of these point charges we obtain $\left(z_{02}/z_{01}\right)^{l_{1}-l_{2}}.$ This result does not depend on $q_1$ and $q_2$ and if we know magnitudes of two $l$ components of the electric fields of two point charges, we can deduce the ratio between the point charge locations.

\section{Results}

\begin{figure}
\includegraphics[width=8cm]{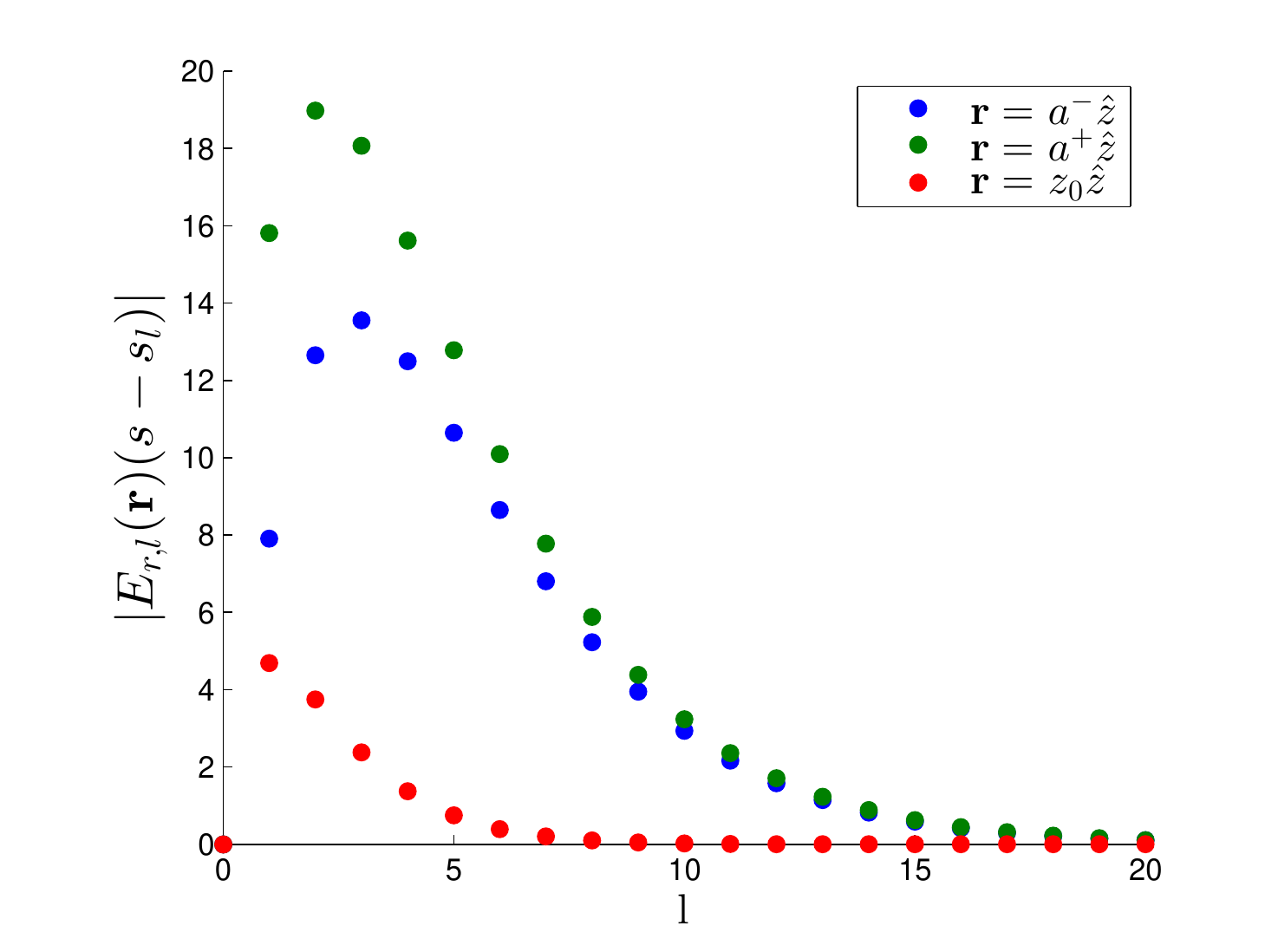}
\protect\caption{\label{fig:cont_combined_1}$|E_{r,l}(\mathbf{r})(s-s_l)|$ at $\mathbf{r}=a^{-}\hat{\mathbf{z}}$ and $\mathbf{r}=a^{+}\hat{\mathbf{z}},$ i.e. just inside and just outside the sphere, and at $\mathbf{r}=z_0\hat{\mathbf{z}}$ as a function of $l$ for $\epsilon_{2}=1,\,z_{0}=1.5a,\, a=30\mathrm{nm}$ }
\end{figure}

We first considered $\epsilon_{2}=1$ and a point charge located at $z_{0}=1.5a,$ where $a=30\mathrm{nm}.$  
In order to exclude the effect of the choice of physical $s$ on the results we decomposed each term in the sum in Eq.\ (\ref{eq:expansion_E}) into $\left(-4\pi q/\epsilon_{2}\right)s_{l}^{2}\psi_{lm}^{*}\left(\mathbf{r}_{0}\right)\nabla\psi_{lm}\left(\mathbf{r}\right),$ which does not depend on the choice of $s,$ and $1/(s-s_l).$ The size of the last factor is determined by the distance between the physical $s$ and the eigenvalue $s_l.$
We calculated 
\begin{align}
&|E_{r,l}(\mathbf{r})(s-s_l)|=\nonumber \\
& \left|\frac{4\pi q}{\epsilon_{2}}s_{l}^{2}\psi_{l,m=0}^{*}\left(\mathbf{r}_0\right)\frac{\partial \psi_{l,m=0}\left(\mathbf{r}\right) }{\partial r}\right|
\end{align} at $\mathbf{r}=a^{-}\hat{\mathbf{z}},$ i.e., just inside the sphere, up to $l=20.$ Note that the spectral components of $\mathbf{E}_0=-\nabla \psi_0$ can be included in the calcuation of $E_{r,l}(\mathbf{r})$ both inside the sphere and at  the sphere surface. We found that the $l=3$ mode with $s_{l}= 0.4286,\epsilon_{1,l=3}=-4/3$ is the most dominant one. In Fig. \ref{fig:cont_combined_1} we present the results as a function of $l.$

We then chose  $\epsilon_{1}= -1.3256, s=0.43$  which are close to the $l=3$ mode resonance. We calculated the electric field for these $s$ and $\epsilon_1 $ values. 
 The calculation of the electric field was performed analytically using Eq.\ (\ref{eq:electric_field}). In Fig. \ref{fig:E_3rd} we present the intensity of the electric field.
%\begin{figure}
%\includegraphics[width=8cm]{potential_3rd_mode}
%\protect\caption{\label{fig:pot_3rd}$\psi$ for a point charge at $z_{0}=1.5a,\, a=30\mathrm{nm},\, s=0.43,\,\epsilon_{2}=1,\,\epsilon_{1}= -1.3256$}
%\end{figure}
\begin{figure}
\label{fig:E_3rd}
\includegraphics[width=8cm]{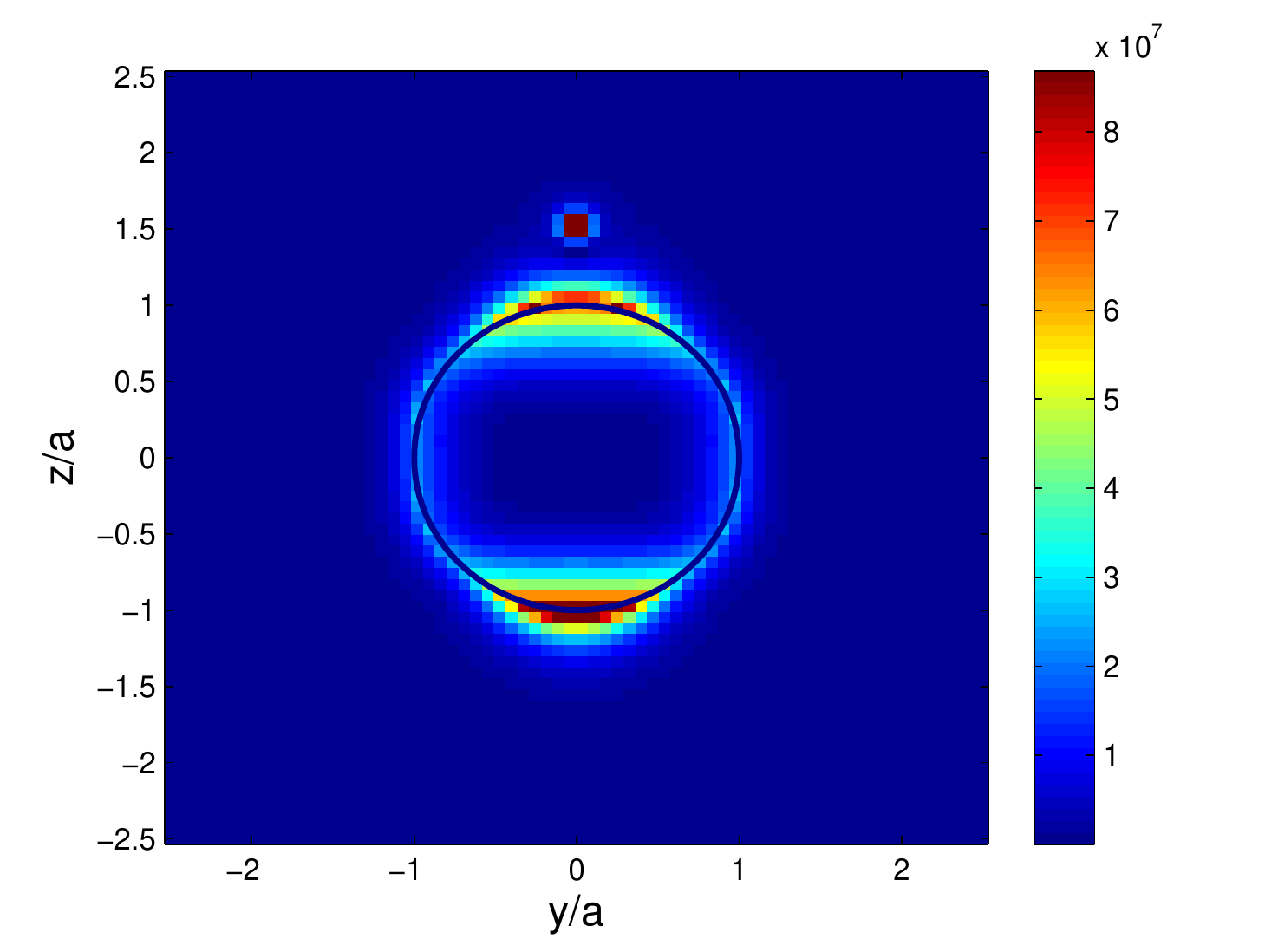}
\protect\caption{\label{fig:E_3rd}$\left|\mathbf{E}\right|^{2}$ for a point charge at $z_{0}=1.5a,\, a=30\mathrm{nm},\, s=0.43,\,\epsilon_{2}=1,\,\epsilon_{1}= -1.3256$}
\end{figure}
It can be seen that the electric field is significantly enhanced with maximal intensity at the interface between the sphere and the host medium at $\theta=0,\pi.$
  
We also calculated $|E_{r,l}(\mathbf{r})(s-s_l)|$ at $\mathbf{r}=a^+\hat{\mathbf{z}},$ i.e. just \emph{outside} the sphere, and at the point charge location $(\mathbf{r}=1.5a\hat{\mathbf{z}}$  --- see Fig.\ \ref{fig:cont_combined_1}).
%We present these results in Figs. \ref{fig:contoutside} and \ref{fig:contz0}. 
The most dominant modes at $\mathbf{r}=a^+\hat{\mathbf{z}}$ and at the point charge location $(\mathbf{r}=1.5a\hat{\mathbf{z}})$ are $l=2$ and $l=1$ respectively. The contributions to the electric field inside and outside the sphere do not need to have the same $l$ dependence since continuity of $D_r$ for each mode is satisfied for the eigenvalue $\epsilon_{1l}$ but not for $\epsilon_1.$ 
%\begin{figure}
%\includegraphics[width=8cm]{cont_outside}
%\protect\caption{\label{fig:contoutside} $|E_{r,l}(\mathbf{r}=a\hat{z})(s-s_l)|$ outside the sphere as a function of $l$ for $\epsilon_{2}=1$ and a point charge located at $z_{0}=1.5a,\, a=30\mathrm{nm}$ }
%\end{figure}
%\begin{figure}
%\includegraphics[width=8cm]{cont_z0}
%\protect\caption{\label{fig:contz0} $|E_{r,l}(\mathbf{r}=\mathbf{r}_0)(s-s_l)|$ as a function of $l$ for $\epsilon_{2}=1$ and a point charge located at $z_{0}=1.5a,\, a=30\mathrm{nm}$ }
%\end{figure}
 
We then calculated $|E_{r,l}(\mathbf{r})(s-s_l)|$ for $z_0=2a.$ The most dominant modes of the electric field at $\mathbf{r}=a^{-}\hat{\mathbf{z}},$ $\mathbf{r}=a^+\hat{\mathbf{z}}$ and at the point charge location $(\mathbf{r}=2a\hat{\mathbf{z}})$  were found to be $l=2,l=1,$ and $l=1$ respectively.
%  with $\epsilon_{1l}=-1.5,\, s_l=0.4$. We then chose  $\epsilon_{1}=-1.4691, s=0.405$  which are close to the $l=2$ mode resonance. 
%We then performed the calculations for $z_0=2a.$ The most dominant mode for the electric field at $\mathbf{r}=a^{-}\hat{z}$ was found to be $l=2$  with $\epsilon_{1l}=-1.5,\, s_l=0.4$. We then chose  $\epsilon_{1}=-1.4691, s=0.405$  which are close to the $l=2$ mode resonance. 
%In Figs. \ref{potential_2nd_mode}, \ref{E_2nd_mode} we present the potential and the intensity of the electric
%field. It can be seen that the electric field is, again, enhanced with maximal intensity at the interface between the sphere and the host medium at $\theta=0,\pi.$ The electric field was less concentrated spatially compared to the excitation of the $l=3$ mode as expected.
%
%%\begin{figure}
%%\includegraphics[width=8cm]{potential_2nd_mode}
%%\protect\caption{\label{potential_2nd_mode}$\psi$ for a point charge at $z_{0}=2a,\, a=30\mathrm{nm},\, s=0.405,\,\epsilon_{2}=1,\,\epsilon_{1}= -1.4691.$ }
%%\end{figure}
%\begin{figure}
%\includegraphics[width=8cm]{electric_field_2nd_mode}
%\protect\caption{ \label{E_2nd_mode}$\left|\mathbf{E}\right|^{2}$ for a point charge at $z_{0}=2a,\, a=30\mathrm{nm},\, s=0.405,\,\epsilon_{2}=1,\,\epsilon_{1}=-1.4691$}
%\end{figure}
%We also calculated $|E_{r,l}(\mathbf{r})(s-s_l)|$ at $\mathbf{r}=a^+\hat{z}$ and at the point charge location $(\mathbf{r}=2a\hat{z})$ and the most dominant mode was $l=1$ in both cases.

\begin{figure}
\includegraphics[width=8cm]{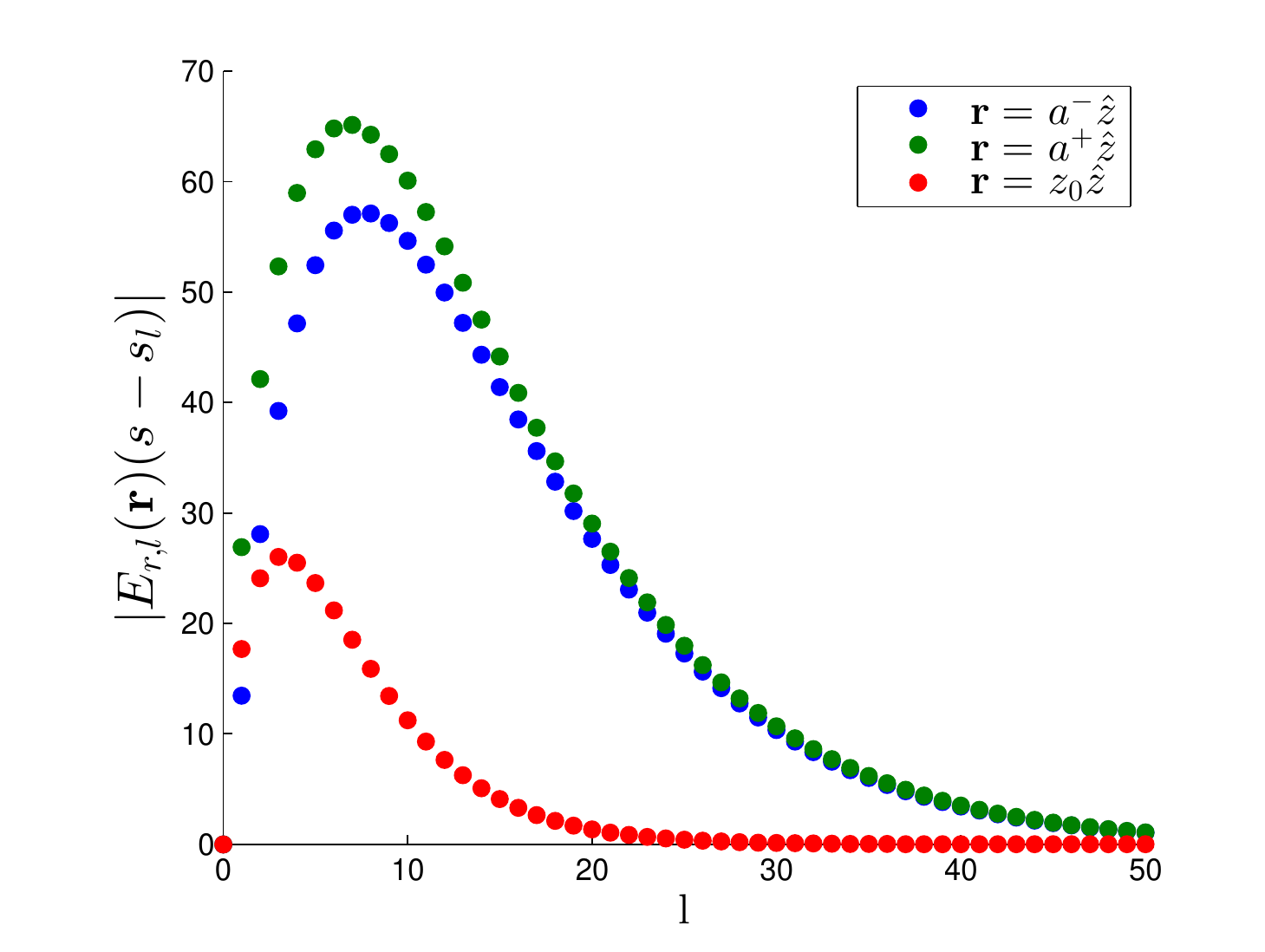}
\protect\caption{\label{fig:cont_combined_3}$|E_{r,l}(\mathbf{r})(s-s_l)|$ at $\mathbf{r}=a\hat{\mathbf{z}}$ inside and outside the sphere and at $\mathbf{r}=z_0\hat{\mathbf{z}}$ as a function of $l$ for $\epsilon_{2}=1,$ a point charge located at $z_{0}=1.15a,\, a=30\mathrm{nm}$ }
\end{figure}

\begin{figure}
\includegraphics[width=8cm]{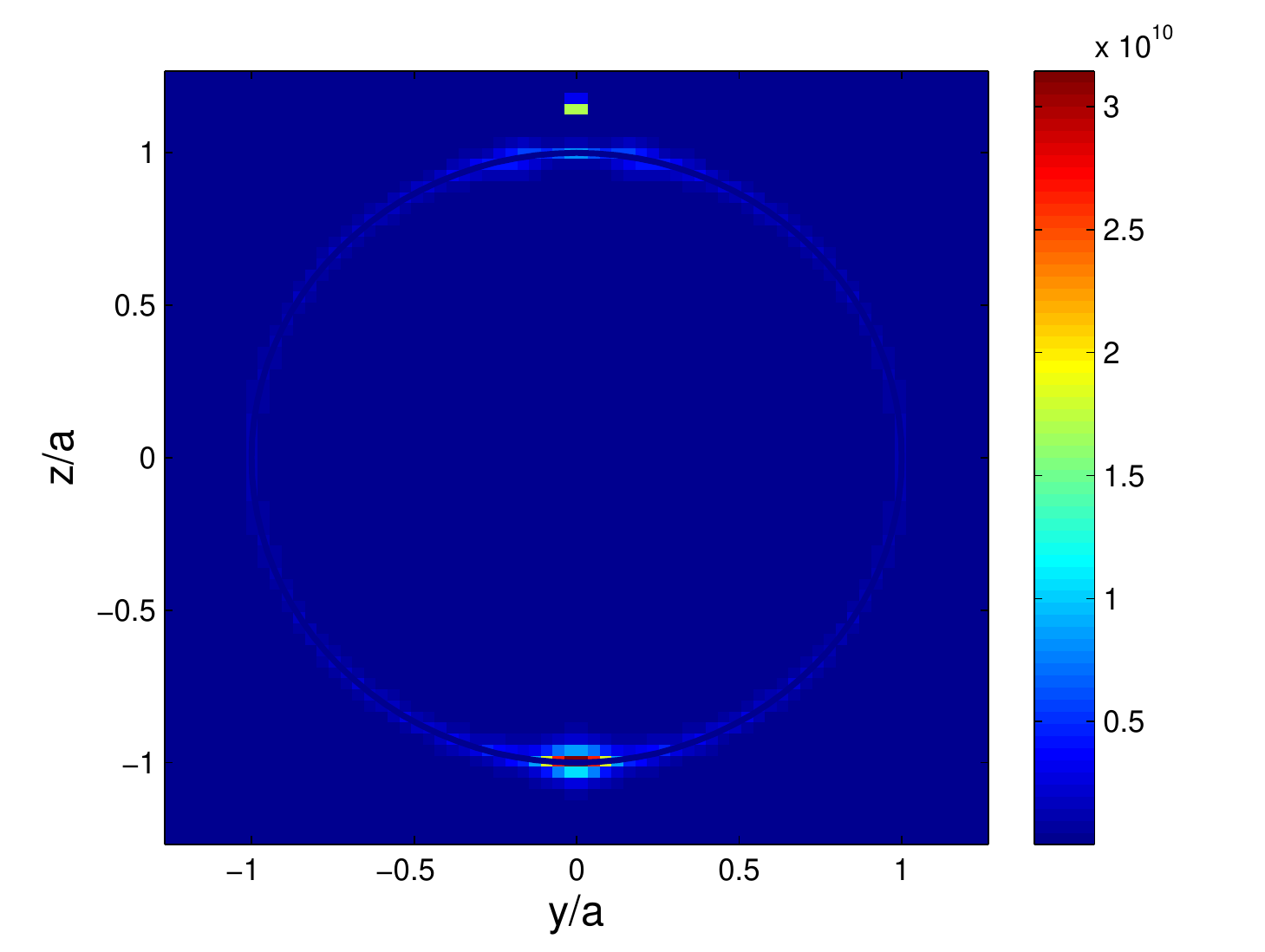}
\protect\caption{ \label{E_3}$\left|\mathbf{E}\right|^{2}$ for a point charge at $z_{0}=1.15a,\, a=30\mathrm{nm},\, s=0.487,\,\epsilon_{2}=1,\,\epsilon_{1}=-1.0534$}
\end{figure}

Then, for a point charge located at $\mathbf{r}_0=1.15a\hat{\mathbf{z}}$ we calculated $|E_{r,l}(\mathbf{r})(s-s_l)|$ at both $\mathbf{r}=a^{-}\hat{\mathbf{z}}$ and $\mathbf{r}=a^{+}\hat{\mathbf{z}}$ and at the point charge location $(\mathbf{r}=1.15a\hat{\mathbf{z}}),$ and the most dominant modes were found to be $l=8,l=7,$ and $l=3,$ respectively (see Fig.\ \ref{fig:cont_combined_3}). Thus, as the point charge approaches the sphere interface the most dominant modes are of higher order, including for a measurement at the point charge location. These calculations necessitated 50 modes in the expansion. Here, we were interested to excite a high order mode and compromise on intensity, which is high anyway. We therefore chose $s=0.487$ which corresponds to $\epsilon_1=-1.0534$ and is close to  the $s_{l=20}= 0.4878$ resonance. In Fig. \ref{E_3} we present the electric field intensity. It can be seen that the field intensity is greatly enhanced. In addition, the field is highly localized at $\theta=\pi$ with $\exp(-1/2)$ of the maximal intensity at $\simeq 2\mathrm{nm}$ from the maximum. 

%\begin{figure}
%\includegraphics[width=8cm]{potential_3}
%\protect\caption{\label{potential_3}$\psi$ for a point charge at $z_{0}=1.15a,\, a=30\mathrm{nm},\, s=0.487,\,\epsilon_{2}=1,\,\epsilon_{1}= -1.0534.$ }
%\end{figure}

\begin{figure}
\includegraphics[width=8cm]{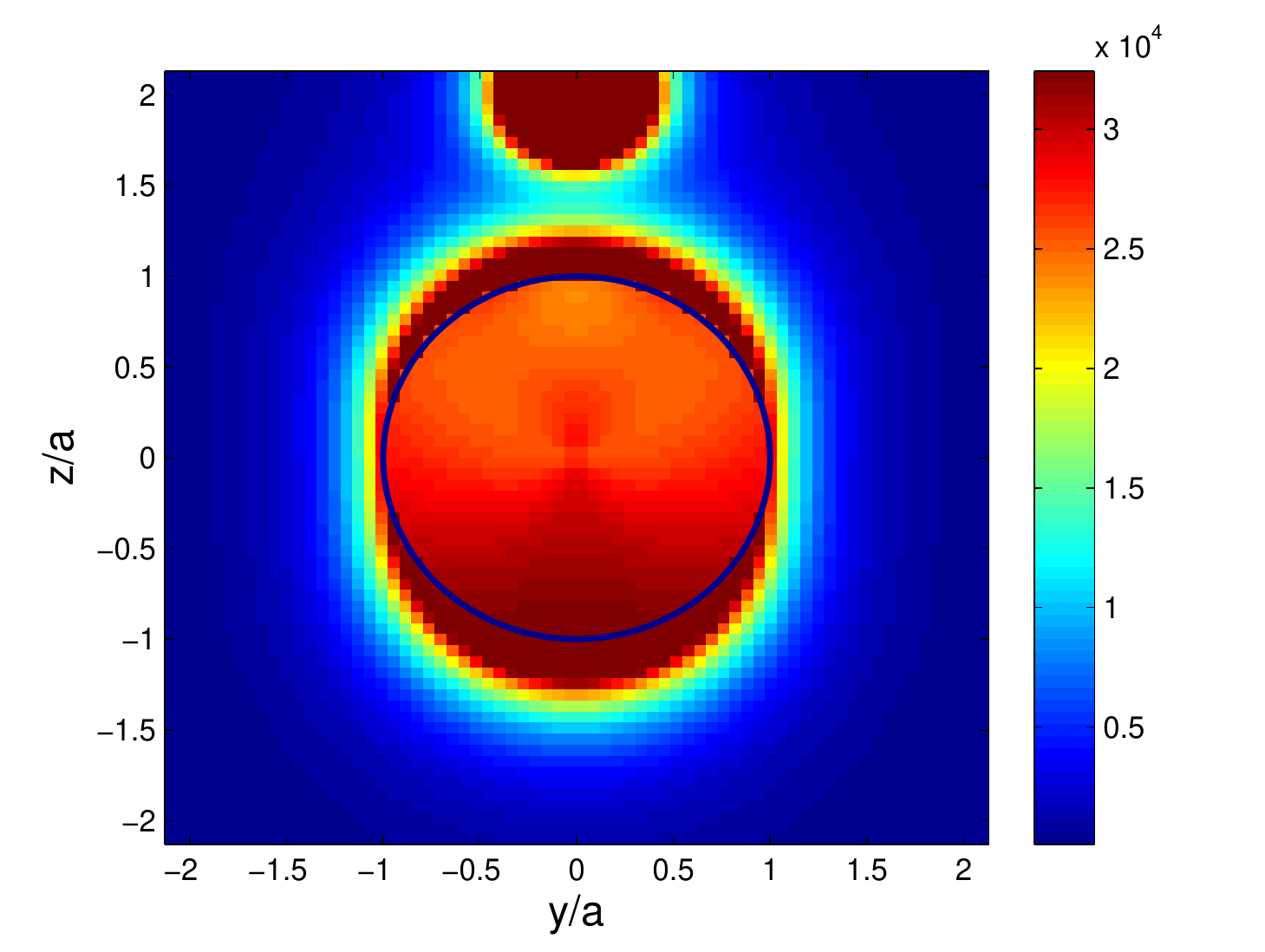}
\protect\caption{\label{fig:intensity_1st_mode}$\left|\mathbf{E}\right|^{2}$ for a point charge at $z_{0}=2a,\, a=30\mathrm{nm},\, s=1/3,\,\epsilon_{2}=1.69+0.08i,\,\epsilon_1=-3.38+0.192i$ }
\end{figure}

We were then interested to consider a system that is close the $l=1$ resonance. The electric field of the $l=1$ mode extends far from the interface and does not vanish at the origin. A resonance of this mode occurs when the material parameters satisfy $\epsilon_2\approx -\epsilon_1/2$ and with small and positive $\mathrm{Im}(\epsilon_1)$ and $\mathrm{Im}(\epsilon_2)$ we can approach this resonance. We chose $\epsilon_1=-3.38+0.192i$ (silver at 380nm) and $\epsilon_2= 1.69+0.08i$ and placed a point charge at $\mathrm{r}_0=2a\hat{\mathbf{z}}$. In Fig.\ \ref{fig:intensity_1st_mode} we present $\left|\mathbf{E}\right|^{2}$ in space. It can be seen that there is a strong electric field inside the sphere even thought it is a conductor. 

\begin{figure}
\includegraphics[width=8cm]{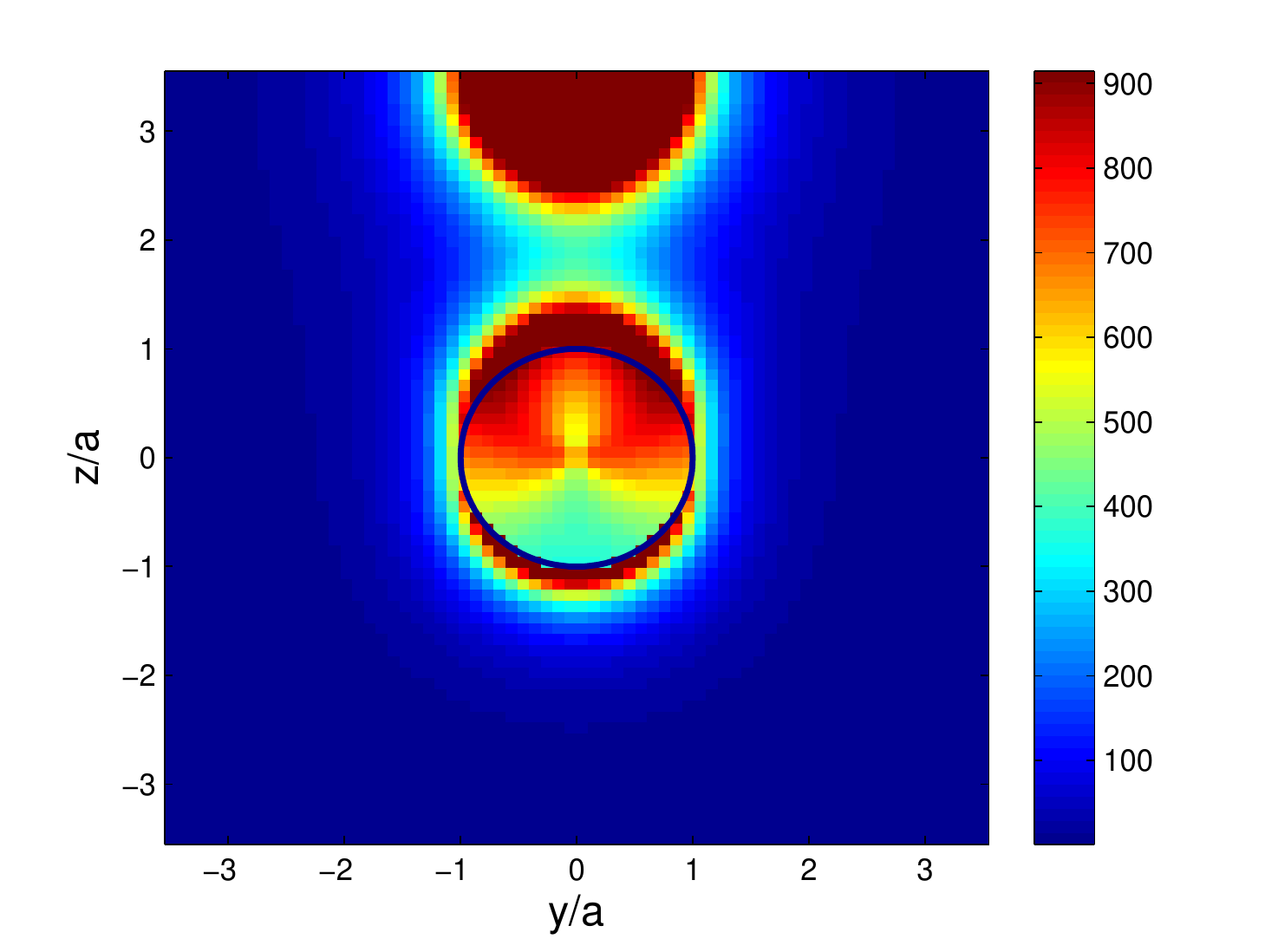}
\protect\caption{\label{fig:intensity_s_0_3}$\left|\mathbf{E}\right|^{2}$ for a point charge at $z_{0}=3.5a,\, a=30\mathrm{nm},\, s=0.3,\,\epsilon_{2}=1,\,\epsilon_1=-2.33$ }
\end{figure}

\begin{figure}
\includegraphics[width=8cm]{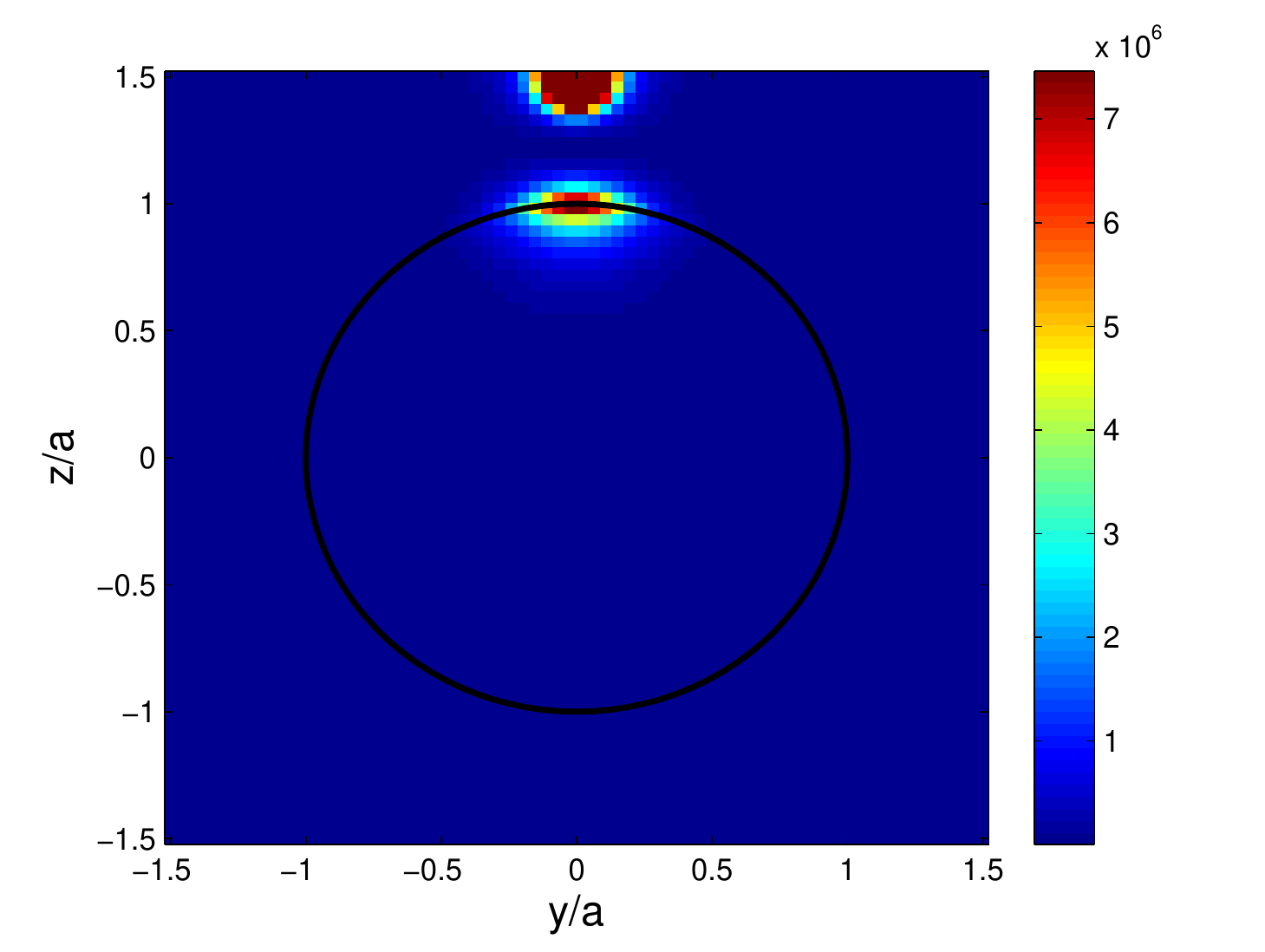}
\protect\caption{\label{fig:intensity_s_0_492}$\left|\mathbf{E}\right|^{2}$ for a point charge at $z_{0}=1.5a,\, a=30\mathrm{nm},\, s=0.492,\,\epsilon_{2}=1,\,\epsilon_1= -1.0325$ }
\end{figure}

Finally, we calculated $\left|\mathbf{E}\right|^{2}$ for setups in which $s$ is smaller or larger than $s_l$ of all the dominant modes. In these setups the low and high order modes interfere constructively at $\theta=0.$ In Fig. \ref{fig:intensity_s_0_3} we present $\left|\mathbf{E}\right|^{2}$ for $z_{0}=3.5a$ and $s=0.3$ $\left(\epsilon_{2}=1,\,\epsilon_1=-2.33\right)$ which is smaller than all the eigenvalues $s_l.$ It can be seen that the intensity is strong at $\theta=0$ and that the electric field extends far from the sphere surface since $s$ is closer to $s_l$ of the low order modes. In Fig. \ref{fig:intensity_s_0_492} we present $\left|\mathbf{E}\right|^{2}$ for $z_{0}=1.5a$ and $s=0.492$ $\left(\epsilon_{2}=1,\,\epsilon_1= -1.0325\right)$ which is larger than the eigenvalues $s_l$ of the dominant modes. The intensity is again strong at $\theta=0$ and is spatially concentrated since $s$ is closer to $s_l$ of the high order modes which are associated with high spatial frequencies.

To verify our results we checked the continuity of the physical $D_r$ at the interface. In the electric field expansions $\mathbf{E}_{0}$ is continuous and all the eigenstates satisfy continuity of $D_r$ with their $\epsilon_{1l}$ eigenvalue. Thus, none of the terms in the expansion is expected to satisfy continuity of physical $D_r$ at the interface. Our calculations showed that the physical $D_r$ is continuous at the interface for various $s$ values to a high accuracy. 

The calculations were performed using a grid of $70\times70$ in $y,z$ axes. In our calculations of the potential and the electric field in all space, the running times on a single core were $\sim2,3,8\mathrm{s}$ for $z_0=2a,1.5a,1.15a,$ respectively, which can be reduced by an order of magnitude with code optimization in Matlab.

\section{Discussion}

We presented an analytic expansion of the potential and the electric field for a setup of an $\epsilon_1$ sphere embedded in an $\epsilon_2$ host medium, where the permittivity values of the sphere and the host medium can take any value. For a point charge on the $z$ axis at $z_0$ the expansion only includes the $m=0$ terms and involves up to 20 terms when $z_0 \gtrsim 1.5a.$ For a given charge location and measurement point a dominant mode can be readily identified and one can select a sphere permittivity value which is close to the mode resonance in order to obtain a significant enhancement of the electric field. 

We placed a point charge at $z_0=1.5a, z_0=1.15a,z_0=2a$ and $z_0=3.5a$ and permittivity values which are close to a resonance. We observed very high enhancement of the electric field. Interestingly, a significant electric field can exist inside the sphere even it is a conductor, when $\epsilon_1/\epsilon_2$ is close to $\left(\epsilon_{1}/\epsilon_2\right)_l$ of a dominant mode. The high order modes become non-negligible as the point charge approaches the sphere surface. The low order modes decay more slowly and generate an electric field away from the surface. Very high resolution is obtained when a high order mode is excited since high order modes are associated with high spatial frequencies. When $s\approx (s_l+s_{l+1})/2$ the dominant $l$ and $l+1$ modes interfere constructively at $\theta=\pi.$
When $\epsilon_{1}\gtrsim-2\epsilon_{2},$ the radial field component of all the modes at $\theta=0$ interfere constructively and generate a strong signal dominated by the low order modes which extend far from the sphere surface.
 Similarly, when $\epsilon_{1}\lesssim-\epsilon_{2}$ the radial field component of all the modes at $\theta=0$ interfere constructively and a strong signal dominated by the high order modes which are associated with high spatial frequencies is generated.
%Moreover, we showed that when scaling down the system by a factor $b$ the intensity of the scattered electric field scales up by $b^2.$

We showed that the spectral information at the sphere surface can be utilized to calculate the point charge location without knowing its magnitude. In addition, when the system is close to a resonance the high order modes of the electric field can be retrieved. These may have relevance for near field imaging of objects that are not at the surface. To assist in balancing the magnitudes of the signals from distanced sources, the magnitude of the light sources can be larger for larger $r\mathrm{s},$ which can be achieved by back light. Gain can both enhance the incoming field and enable $s$ that is closer to the $s_l$ resonances which are real. Another possible mechanism to enable detection of high order modes in the expansion of the electric field of a point charge that is not very close to the surface is to mediate them through resonant particles inside the medium which enhance them, similarly to the isolated sphere. Since we can calculate the point charge location for a single point charge, selectively exciting local points which radiate at different times may enable to retrieve their locations too \cite{hanne2015sted,fernandez2008fluorescent}.  A similar analysis can be formulated for a setup of a flat slab in a host medium \cite{Farhi2014exact} where the spectrum of the eigenvalues is continuous. 
  
% Techniques such as STED and PALM enable to  . 
%
%
%If the point charges are localized in a thin layer with $r>a$ then their magnitude may be detected.  In addition, if $s$ is not too close to a resonance, the field intensity is generated by several modes and there is more spectral information. This may assist in retrieving point charge distances and magnitudes since each charge distance has a spectral fingerprint . In order to generate more spectral information one can also vary $s$ and measure the field intensity. Lastly, fields on the sphere envelope and at different radii may be also measured to retrieve point charge locations. 
 Potential applications are enhancement of spontaneous emission of a molecule by an antenna \cite{eggleston2015optical}, where the point charge and the sphere can model the molecule and the antenna, respectively, sensing, modeling a tip in proximity to a metallic nanosphere, near field imaging, and Raman spectroscopy. Finally, since the expansion employs a small number of terms for a single point charge source, calculating the potential and the electric field in all space is very fast. 
 
\bibliographystyle{apsrev}

\bibliography{bib2}

\end{document}